\documentstyle[12pt]{article}
\newcommand\btd{\raise 2pt \hbox{$\hat\bigtriangledown$}\hskip 1.5pt}
\newcommand\bt{\raise 2pt \hbox{$\bigtriangledown$}\hskip 1.5pt}
\begin{document}

\title{\large\bf New Darboux Transformation for Hirota-Satsuma coupled
KdV System}
\author{Heng Chun Hu and Q. P. Liu \\[.2cm]
China University of
Mining and Technology,\\
 Beijing 100083, China.\\
International Center for Theoretical Physics,\\
34014 Trieste, Italy.}
\date{}
\maketitle

\begin{abstract}
A new Darboux transformation is presented for the Hirota-Satsuma
coupled KdV system. It is shown that this Darboux transformation
can be constructed by means of  two methods: Painlev\'{e} analysis
and  reduction of a binary Darboux transformation. By iteration of
the Darboux transformation, the Grammian type solutions are found
for the coupled KdV system.
\end{abstract}
\newpage

\section{Introduction}
Hirota and Satsuma \cite{hs1} proposed the very first coupled KdV system,
which describes interactions of two long
waves with different dispersion relations.
They then constructed the three-soliton solutions and five conserved
quantities for this system. In a following paper \cite{hs2}, these authors
shown that the coupled KdV system is the four-reduction of the celebrated KP
hierarchy  and its soliton solutions can be derived from those of the KP
equation.

It is known that a solitonic equation normally possesses a Lax pair. Using
the Wahlquist-Estabrook prolongation procedure, Dodd and Fordy \cite{df}
found a Lax
representation for the Hirota-Satsuma coupled KdV (HS-KdV) system. Meanwhile,
Wilson \cite{w} observed that the HS-KdV system is just an example of many
integrable systems arose from the Drinfeld-Sokolov theory. A B\"{a}cklund
transformation is constructed for this system by Levi \cite{le}. Very
recently, some new solutions for HS-KdV system are presented \cite{ta}.

Most integrable equations have Darboux Transformations (DT),
which are very effective to construct solutions. We notice that Leble and
Ustinov \cite{lu} found a DT for the HS-KdV system. They started with an
elementary DT for a more general spectral problem, then they found that a
proper reduction led to a DT for HS-KdV system. In general, an integrable
system may possess more than one DT. For example, the KdV equation has classical
DT and binary DT as well. We are interested in finding a new DT for the HS-KdV
system. We will show that this new DT appears in two ways,  from
Painlev\'{e} analysis and from reduction of the binary DT for a more
general spectral problem (we refer to  \cite{li}-\cite{eg} and the
references there for Painlev\'{e} analysis approach to DT).

The paper is set out as follows. In section 2 we construct a new
DT for the HS-KdV system within the framework of Painlev\'{e}
analysis. In section 3, we show that a DT can be obtained from the
reduction of a general DT, then we prove that this DT is
equivalent to the one found in section  2. In section 4 we discuss
the iteration of the DT and show that the solutions for the HS-KdV
system can be represented as Grammians. Final section presents our
conclusion.

\section{DT from Painlev\'{e} analysis}
The HS-KdV system reads
\begin{equation}\label{hs}
u_t=\frac{1}{2}u_{xxx}+3uu_x-6vv_x, \quad
v_t=-v_{xxx}-3uv_x.
\end{equation}
where the subscripts denote partial derivatives.

A detailed Painlev\'{e} analysis of this system was performed by Weiss
\cite{w1}\cite{w2}. He found that this system has the
Painlev\'{e} property. Indeed, there are two branches and one of them,
principal branch, is
 \begin{equation}\label{exp}
\quad u=\tau^{-2}\sum_{j=0}^{\infty}u_j\tau^j, \quad
v=\tau^{-1}\sum_{j=0}^{\infty}v_j\tau^j,
\end{equation}
and the resonances are
\[
j=-1,0,1,4,5,6
\]
By truncating the expansions (\ref{exp}) on the constant level, one has a
 transformation \begin{equation}\label{bt}
u=u_2+2(\ln\tau)_{xx}, \quad v=v_1+\frac{v_0}{\tau}
\end{equation}
where $u_2$ and $v_1$  solve the HS-KdV system (\ref{hs}) and
$\tau$, $u_2$, $v_0$, $v_1$ have to satisfy the following
overdetermined system
\begin{eqnarray*}
&&\tau_t+\tau_{xxx}+3\tau_xu_2=2\tau_{x}\vartheta,\\
&&v_0=\tau_xH,\\
&&{\tau_t\over \tau_x}-{1\over 2}\{\tau;x\}={3\over 4}H^2+\vartheta,\\
&&\vartheta_{x}^{2}=(\lambda^2+\vartheta^2)H^2,\\
&&v_1=-{v_{0x}\over 2\tau_{x}}-{1\over 3}(\lambda^2+\vartheta^2)^{1\over 2},\\
&&H_t=-\left[H_{xx}+{1\over 4}H^3+\vartheta H+{2\over 3}\{\tau;x\}H\right]_x,
\end{eqnarray*}
where $\{\tau;x\}=\left({\tau_{xx}\over\tau_x}\right)_x-{1\over2}\left({\tau_{xx}\over\tau_x}\right)^2$
is the Schwarzian derivative.

To obtain a Lax pair for the HS-KdV system (\ref{hs}), Weiss introduced a new
variable
\begin{equation}\label{W}
W={\tau_{xx}\over\tau_x}
\end{equation}
and derived the equations for $H$ and $W$, which are modified
Hirota-Satsuma system. The Miura map between HS-KdV system and its
modification is
\begin{eqnarray*}
u_2&=&-{1\over 2}\left[W_x+{1\over 2}W^2+{1\over
2}H^2-{2\over3}\vartheta\right], \\
 v_1&=&-{1\over 2}\left[H_x+WH+{2\over
3}(\lambda^2+\vartheta^2)^{1/2}\right].
\end{eqnarray*}
Taking
\begin{equation}\label{WH}
W+H=2{\eta_{1x}\over\eta_1},\quad W-H=2{\eta_{2x}\over\eta_{2}},
\end{equation}
\begin{equation}
\vartheta=\lambda\sinh\alpha,\quad \alpha=\ln({\eta_1\over\eta_2}),
\end{equation}
one arrives at the linear problem
\begin{eqnarray}\label{lin-x-lam}
\eta_{1xx}+(u_2+v_1)\eta_1&=&-\lambda\eta_2,\\
\eta_{2xx}+(u_2-v_1)\eta_2&=&\lambda\eta_1,
\end{eqnarray}
and
\begin{eqnarray}
\eta_{1t}&=&-{1\over
2}(u_{2x}-2v_{1x})\eta_1+(u_2-2v_1)\eta_{1x}-2\lambda\eta_{2x},\\
\eta_{2t}&=&-{1\over
2}(u_{2x}+2v_{1x})\eta_2+(u_2+2v_1)\eta_{2x}+2\lambda\eta_{1x},
\label{lin-t-lam}
\end{eqnarray}
where the parameter $\lambda$ is scaled for convenience. This is
the linear problem found by Dodd and Fordy via the prolongation
approach \cite{df}.

 We intend to construct a DT for the HS-KdV
systems (\ref{hs}). As a first step, we rewrite the transformation
(\ref{bt}) between two solutions of the system (\ref{hs}) in terms
of the solutions of linear problem
(\ref{lin-x-lam}-\ref{lin-t-lam}). Using (\ref{W}) and (\ref{WH}), we obtain
\begin{equation}\label{potential} u-u_2=2(\ln\tau)_{xx},\quad
v-v_1=\frac{\eta_2\eta_{1x}-\eta_{2x}\eta_{1}}{\tau},
\end{equation}
where $(\eta_1,\eta_2) $ is a solution of the linear system
(\ref{lin-x-lam}-\ref{lin-t-lam}).   The $x$-derivative of function $\tau$
is defined in terms of $\eta_1$ and $\eta_2$ as follows
\begin{equation}\label{tau-x} \tau_x=\eta_1\eta_2.
\end{equation}
The transformation (\ref{potential}) is the Darboux transformation on
the potential level.
Next, we need to find the transformations for the wave function.
Let us suppose that $(\phi_1, \phi_2)$ solves the linear systems
(\ref{lin-x-lam}-\ref{lin-t-lam}) with $\lambda=\mu$.
For the new wave function, we make the following Ans\"{a}tz
\begin{equation}\label{newave}
\hat{\phi}_1=\phi_1+{f\over\tau}, \quad \hat{\phi}_2=\phi_2+{g\over\tau},
\end{equation}
where $f=f(x,t)$ and $g=g(x,t)$ are the functions to be determined.
By requiring the new wave function $(\hat{\phi}_1,\hat{\phi}_2)$ solves
the linear problem (\ref{lin-x-lam}-\ref{lin-t-lam}) with replacement $u_2 \to u$,
$v_1 \to v$ and $\lambda\to\mu$, we are led to two equations
\begin{eqnarray}
&&\phi_{1xx}+(u_2+v_1)\phi_1+\mu\phi_2+\nonumber\\
&&\quad+{1\over\tau}\Big[f_{xx}+(u_2+v_1)f+2\tau_{xx}\phi_1+
(\eta_{1x}\eta_2-\eta_1\eta_{2x})\phi_1+\mu
g\Big]+\nonumber\\
&&
\quad\quad+{1\over\tau^2}\Big[-2f_x\tau_x+f\tau_{xx}-2\tau_{x}^2\phi_1
+(\eta_{1x}\eta_2-\eta_1\eta_{2x})f \Big]=0,
\label{first}
\end{eqnarray}
and
\begin{eqnarray}
&&\phi_{2xx}+(u_2-v_1)\phi_2-\mu\phi_1+\nonumber\\
&&\quad+{1\over\tau}\Big[g_{xx}+(u_2-v_1)g+2\tau_{xx}\phi_2
-(\eta_{1x}\eta_2-\eta_1\eta_{2x})\phi_2-\mu
f\Big]+\nonumber\\
&&\quad\quad+{1\over\tau^2}\Big[-2g_x\tau_x+g\tau_{xx}-
2\tau_{x}^2\phi_2-(\eta_{1x}\eta_2-\eta_1\eta_{2x})g \Big]=0.
\label{second}
\end{eqnarray}
Let us consider the equation (\ref{first}) first.
The first part of this equation, the coefficient of $\tau^0$, is identically
zero.
Equating the coefficients of $\tau^{-1}$ and $\tau^{-2}$ to zero respectively,
we have
\begin{eqnarray}
f_{xx}+(u_2+v_1)f+2\tau_{xx}\phi_1+(\eta_{1}\eta_2-\eta_1\eta_{2x})\phi_1+\mu
g & = & 0 \label{f1},\\
-2f_x\tau_x+f\tau_{xx}-2\tau_{x}^2\phi_1+(\eta_{1x}\eta_2-\eta_1\eta_{2x})f&=&0.
\label{f2}
\end{eqnarray}
From (\ref{f2}), we obtain
\begin{equation}\label{f3}
f_x-f{\eta_{1x}\over\eta_1}+\eta_1\eta_2\phi_1=0.
\end{equation}
Now, using (\ref{f1}) and ({\ref{f3}), we have
\begin{equation}\label{f}
\lambda\eta_2 f-\mu\eta_1 g=\eta_1\eta_2(\phi_1\eta_{1x}-\phi_{1x}\eta_1).
\end{equation}
A similar consideration for the other equation (\ref{second})
 provides us
\begin{equation}\label{g}
-\mu\eta_2 f+\lambda\eta_1 g=\eta_1\eta_2(\eta_2\phi_{2x}-\eta_{2x}\phi_2).
\end{equation}
In this way, we obtain a system of linear equations for $f$ and $g$,
namely, (\ref{f}) and (\ref{g}). Solving this linear system, we get the
expressions for $f$ and $g$ as follows  \begin{eqnarray}\label{g1}
&&f={\eta_1\over
\lambda^2-\mu^2}\Big[\mu(\eta_2\phi_{2x}-\eta_{2x}\phi_2)+\lambda(\phi_1\eta_{1x}-\phi_{1x}\eta_{1})\Big],\\
\label{g2}
&&g= {\eta_2\over
\lambda^2-\mu^2}\Big[\mu(\phi_1\eta_{1x}-\phi_{1x}\eta_{1})+\lambda(\eta_2\phi_{2x}-\eta_{2x}\phi_{2})\Big].
\end{eqnarray}
Thus, we obtain the explicit expressions for the transformations for the wave
functions, that is, the equations (\ref{newave}) with $f$ and $g$
given by (\ref{g1}-\ref{g2}). However, this is not the end of story: we have
to make sure that the transformed wave functions and fields also satisfy the
temporal part of the linear problem. But first we need to obtain the time
evolution of $\tau$, which is
\begin{equation}\label{tau-t}
\tau_t=2\lambda(\eta_1^2-\eta_2^2)-2\eta_{1x}\eta_{2x}-u_2\eta_1\eta_2.
\end{equation}
Verifying that the new wave functions and new fields obtained under the
transformation are indeed the solutions for transformed linear systems is too
tedious to perform by pen and paper, but we do check its validity by means of
MAPLE. Thus, we obtain a new DT for the HS-KdV system (\ref{hs}). We summarize our
results as

{\bf Proposition} {\it Let $(\eta_1,\eta_2)$ be the
solutions of the linear systems (\ref{lin-x-lam}-\ref{lin-t-lam}) and
 $(\phi_1,\phi_2)$ be the solutions of the same linear systems with
 the spectral parameter $\mu$. Let the function  $\tau$ be
defined by equations (\ref{tau-x}) and (\ref{tau-t}). Then the new
field variables and the new wave functions defined via (\ref{potential}) and
(\ref{newave}) with $f$ and $g$ given by (\ref{g1}) and (\ref{g2}) solve the
linear system (\ref{lin-x-lam}-\ref{lin-t-lam}) with $\mu$ as the spectral
parameter}.


\section{DT from reduction}
Our DT constructed above is different from the known DT obtained
by Leble and Ustinov \cite{lu}. Indeed, their DT is of kind
classical DT while our one is of kind binary DT: we used both the
wave functions and the adjoint wave functions in our construction.
Furthermore, our approach is based on Painlev\'e analysis while
they obtain their DT by means of reductions. Noticing these
differences, we are interested in constructing a DT for (\ref{hs})
by reducing the binary DT for a general linear problem.

As in \cite{lu}, we consider the following linear problem
\begin{equation}\label{li}
(\partial^2+F\partial+U)\Psi=\mu\sigma_3\Psi,
\end{equation}
where $\mu$ is the spectral parameter and
\begin{equation}\label{fu}
F=\left(\begin{array}{cc}f_{11}&f_{12}\\
f_{21}&f_{22}\end{array}\right),\quad
U=\left(\begin{array}{cc}u_{11}&u_{12}\\
u_{21}&u_{22}\end{array}\right),\quad \sigma_3=\left(\begin{array}{cc}1&0\\
0&-1\end{array}\right).
\end{equation}
If
\begin{equation}\label{reduc}
F=0, \quad u_{11}=u_{22}=u, \quad u_{12}=u_{21}=v,
\end{equation} above linear problem is
equivalent to the spatial part of the linear problem for the
HS-KdV system given in last section.

The binary Darboux transformation has been constructed in general
case, for example,  \cite{ni} contains a detailed discussion on this
issue. In the present case, we consider
\begin{equation}\label{Lin}
L\Theta\equiv\Big[\partial_y-\sigma_3(\partial^2+F\partial+U)\Big]\Theta=0,
\end{equation}
where $y$ is a new independent variable, $F$ and $U$ are those matrices
given by (\ref{fu}). It is easy to see that (\ref{li}) is a dimension
reduction of (\ref{Lin}).

What we will do next is to study reduction of the general binary DT and
construct a proper DT for HS-KdV system (\ref{hs}). To construct a binary
DT.
we also need to consider the adjoint linear problem
\begin{equation}\label{LinA}
\Big[-\partial_y-(\partial^2-\partial F^t+U^t)\sigma_3\Big]\Phi=0.
\end{equation}
Now we take $\Theta$ as a matrix solution of the linear system
(\ref{Lin}) and
$\Phi$ as a matrix solution of the adjoint linear
system (\ref{LinA}). Then we define a potential
matrix by
\[
\Omega_x=\Phi^t\Theta,
\]
here and in the sequel superscript $^t$ denotes  matrix transposition.

It is known that
\[
G=1-\Theta\Omega^{-1}\partial^{-1}\Phi^t
\]
satisfies the following equation
\[
(\partial_y-\sigma_3(\partial^2+\hat{F}\partial+\hat{U}))G=G(\partial_y-\sigma_3(\partial^2+F\partial+U)),
\]
where the hatted matrices $\hat{F}$ and $\hat{U}$ are the $2\times 2$
matrices with the new transformed variables.
Therefore, one has a binary DT (for wave functions)
\[
\hat{\Psi}=G\Psi
\]
in this general case. As argued in \cite{ni},
this general DT preserves certain properties of the operator. Therefore, we
choose in the sequel the adjoint wave function as
\[
\Phi=-A\Theta, \quad A=\left(\begin{array}{cc} 0&-1\\ 1&0\end{array}\right).
\]
This reduction is consistent with the DT. Furthermore, a simple calculation
shows that  under this reduction $F$ is not alerted
\[
\hat{F}-F=\Theta\Omega^{-1}\Phi^t-\sigma_3 \Theta\Omega^{-1}\Phi^t\sigma_3=0,
\]
and the potential matrix $\Omega$ is found to satisfy
\begin{equation}\label{om}
\Omega_x=\det(\Theta)A.
\end{equation}
We now have a DT for the linear problem (\ref{Lin}) with $F=0$. Since
the HS-KdV system (\ref{hs}) is a $1+1$-dimensional system, we have to
make further reduction. That is, we need to do dimensional reduction.
Furthermore,
$U$ has to be symmetric and its diagonal entries have to be identical,
namely, $U$ need to take the form (\ref{reduc}).
To this end, we take
\[
\Theta=\left(\begin{array}{cc} \theta_1&\theta_2\\
\theta_2&\theta_1\end{array}\right),
\]
where $(\theta_1,\theta_2)$ is the solution of the following system with
$\mu=\lambda$
\begin{eqnarray*}
&&\left(\begin{array}{cc}\phi_1\\ \phi_2
\end{array}\right)_{xx}+\left(\begin{array}{cc}
u&v\\ v&u\end{array}\right)\left(\begin{array}{cc}\phi_1\\ \phi_2
\end{array}\right)=\mu\sigma_3\left(\begin{array}{cc}\phi_1\\ \phi_2
\end{array}\right),  \\
&&\left(\begin{array}{cc}\phi_1\\
\phi_2\end{array}\right)_t=\left(\begin{array}{cc} -{1\over 2} u_x&v_x\\
v_x&-{1\over 2}u_x \end{array}\right)\left(\begin{array}{cc}\phi_1\\
\phi_2\end{array}\right)+\left(\begin{array}{cc} u+2\mu&-2v\\
-2v&u-2\mu\end{array}\right)\left(\begin{array}{cc}\phi_1\\
\phi_2\end{array}\right)_x.
\end{eqnarray*}
In this case
to show that the general DT is reducible, one has to prove the following
identity holds
 \[
\sigma_3(\partial^2+\hat{U})(1-\Theta\Omega^{-1}\partial^{-1}\Theta A)=
(1-\Theta\Omega^{-1}\partial^{-1}\Theta A)\sigma_3(\partial^2+U),
\]
where $\hat{U}=\left(\begin{array}{cc}\hat{u}&\hat{v}\\ \hat{v}&\hat{u}
\end{array}\right)$. Above equation yields
\begin{equation}\label{U}
\hat{U}-U=(\Theta\Omega^{-1}\Theta A)_x+\sigma_3(\Theta\Omega^{-1}\Theta
A)_x\sigma_3+(\Theta\Omega^{-1})_x\Theta A-\sigma_3(\Theta\Omega^{-1})_x\Theta
A\sigma_3,
\end{equation}
and
\begin{eqnarray}
&&\sigma_3(\Theta_{xx}+U\Theta)\Omega^{-1}\partial^{-1}\Theta
A-\Theta\Omega^{-1}\partial^{-1}(\Theta_{xx}A\sigma_3+\Theta A\sigma_3
U)+\nonumber\\
&& \quad \qquad+\Big[\Theta\Omega^{-1}\Theta_x
A\sigma_3\Theta-\sigma_3\Theta\Omega^{-1}\Theta A\Theta_x
\Big]\Omega^{-1}\partial^{-1}\Theta A=0.
\label{idd}
 \end{eqnarray}
The validity of above identity (\ref{idd}) can be checked by straightforward
calculation. Thus, we have a DT for the linear problems (\ref{li}):
\begin{equation}\label{biwave}
\hat{\Psi}=(1-\Theta\Omega^{-1}\partial^{-1}\Theta A)\Psi,
\end{equation}
and the transformation for $U$ is given by (\ref{U}). This shows
that the general Darboux
transformation can be reduced to the special case: the HS-KdV case. Of course, we
also have to work out the time evolution part of the DT. This can be easily
done by means of MAPLE.

To compare this DT with the DT obtained in last section, we rewrite
above results in terms of
scalars rather than vectors. From the transformation for $U$ (\ref{U}),  we
readily have
\[
\hat{u}-u=2(\ln \rho)_{xx},\quad
\hat{v}-v=2\frac{\theta_1\theta_{2x}-\theta_{1x}\theta_2}{\rho},
\]
where
\[
\rho_x=\theta_1^2-\theta_2^2,
\]
is derived from (\ref{om}). The transformation (\ref{biwave}) for the wave functions
 is
\[
\hat{\phi}_1=\phi_1+{p\over\rho}, \quad \hat{\phi}_2=\phi_2+{q\over\rho},
\]
where
\begin{eqnarray*}
p&=&\frac{\theta_2}{\lambda+\mu}(\theta_2\phi_{1x}-\theta_{2x}\phi_1+\theta_1\phi_{2x}-\theta_{1x}\phi_2)+
\frac{\theta_1}{\lambda-\mu}(\theta_1\phi_{1x}-\theta_{1x}\phi_1+\theta_2\phi_{2x}-\theta_{2x}\phi_2),\\
q&=&\frac{\theta_1}{\lambda+\mu}(\theta_2\phi_{1x}-\theta_{2x}\phi_1+\theta_1\phi_{2x}-\theta_{1x}\phi_2)+
\frac{\theta_2}{\lambda-\mu}(\theta_1\phi_{1x}-\theta_{1x}\phi_1+\theta_2\phi_{2x}-\theta_{2x}\phi_2).
\end{eqnarray*}
For the temporal part, we need time evolution for $\rho$, which is
\[
\rho_t=2\lambda(\theta_1^2+\theta_2^2)+{1\over
2}u(\theta_1^2-\theta_2^2)+\theta_{1x}^2-\theta_{2x}^2.
 \]
With all these in hand, we can show that the transformed variables and wave
functions solve the $t-$part of the linear problem.

Therefore, we do find a DT for the HS-KdV system. The DT presented in this
section seems different from the one obtained in last section, but indeed they
are same. To establish the equivalence, we need to do the following
transformations
\[
\theta_1+\theta_2=f_1,\quad \theta_1-\theta_2=f_2, \quad u_2=u, \quad v_1=v,
\]
then an easy calculation shows that the DT obtained by reduction is nothing
but the one constructed by Painlev\'{e} analysis.


\section{Iterated DT and Exact Solutions}
In general, a DT can be iterated and a compact representation of
solutions can be derived in terms of some special determinants
such as Wronskian, Grammian.  We will show that it is also the
case for our DT.

We notice that our DT for the wave functions (\ref{newave}) can be
written as
\[
\tilde{\phi}_1=\phi_1-\frac{\eta_1}{\tau}\int^x\phi_1\eta_2dx,\qquad
\tilde{\phi}_2=\phi_2-\frac{\eta_2}{\tau}\int^x\eta_1\phi_2dx.
\]
To do iteration, we take $N$ solutions of the linear system
(\ref{lin-x-lam}-\ref{lin-t-lam})  $(\eta_1^{[k]},\eta_2^{[k]})$
 with respect to $\lambda_k$ $(k=1,\dots, N)$. After iterations, we
have the new wave functions
\begin{eqnarray}
\phi_1[N]&=&{1\over \tau[N]}\left|\begin{array}{cccc}
\int^x\eta_1^{[1]}\eta_2^{[1]}dx &\cdots&
\int^x\eta_1^{[1]}\eta_2^{[N]}dx&\eta_1^{[1]}\\
\vdots&\vdots& \vdots&\vdots\\
\int^x\eta_1^{[N]}\eta_2^{[1]}dx&\cdots&\int^x\eta_1^{[N]}\eta_2^{[N]}dx&\eta_1^{[N]}\\[.18cm]
\int^x\phi_1\eta_2^{[1]}dx&\cdots&\int^x\phi_1\eta_2^{[N]}dx&\phi_1\end{array}
\right|,\\[.5cm]
\phi_2[N]&=&{1\over \tau[N]}\left|\begin{array}{cccc}
\int^x\eta_1^{[1]}\eta_2^{[1]}dx &\cdots&
\int^x\eta_1^{[1]}\eta_2^{[N]}dx&\int^x\eta_1^{[1]}\phi_2dx\\
\vdots&\vdots& \vdots&\vdots\\
\int^x\eta_1^{[N]}\eta_2^{[1]}dx&\cdots&\int^x\eta_1^{[N]}\eta_2^{[N]}dx&\int^x\eta_1^{[N]}\phi_2dx\\[.18cm]
\eta_2^{[1]}&\cdots&\eta_2^{[N]}&\phi_2\end{array}\right|,
\end{eqnarray}
and
\[
u[N]=u+2(\ln\tau[N])_{xx}
\]
where
\[
\tau[N]=\det\left(\int^x\eta_1^{[k]}\eta_2^{[\ell]}\right).
\]
Although after a single DT, the field variable $v$ has  a very
nice representation, we do not have a similar one after
iterations. However, if we put $u=w_x$, we may solve the first
equation of (\ref{hs}) and obtain
\[
v^2={1\over 6}(w_{xxx}+3w_x^2-2w_t).
\]
In this way, we can find the transformation for the field $v$.
Therefore, the solutions for (\ref{hs}) are represented in terms
of a Grammian $\tau[N]$.

As a final part of this section, we generate solutions for the
HS-KdV system (\ref{hs}) by means of our DT. Let consider the
simplest case: $u_2=0$, $v_1=c$ ($c$ is an arbitrary constant).
Thus we need to solve
\[
\begin{array}{ll}
\eta_{1xx}=-c\eta_1-i\lambda\eta_2,&
\eta_{2xx}=c\eta_2+i\lambda\eta_1,\\
\eta_{1t}=-2c\eta_{1x}-2i\lambda\eta_{2x},&
\eta_{2t}=2c\eta_{2x}+2i\lambda\eta_{1x},
\end{array}
\]
where for convenience we made a scaling for the spectral
parameter: $\lambda\to i\lambda $ ($i^2=-1$). The general
solutions of the above system read as
\begin{eqnarray}
\eta_1&=&c_1e^{\sqrt{\kappa}(2\kappa t+x)}+c_2
e^{-\sqrt{\kappa}(2\kappa t+x)}+c_3e^{-i\sqrt{\kappa}(-2\kappa
t+x)}
+c_4e^{i\sqrt{\kappa}(-2\kappa t+x)},\label{ex21}\\
\eta_2&=&c_1(c+\kappa)e^{\sqrt{\kappa}(2\kappa t+x)}+c_2(c+\kappa)e^{-\sqrt{\kappa}(2\kappa t+x)}\nonumber\\
&&+ c_3(c-\kappa)e^{-i\sqrt{\kappa}(-2\kappa t+x)}
+c_4(c-\kappa)e^{i\sqrt{\kappa}(-2\kappa t+x)}\label{ex22}
\end{eqnarray}
where $\kappa^2=c^2+\lambda^2 $, $c_j$ $(j=1,\dots,4)$ are
arbitrary constants. Solving
\[
\tau_x=\eta_1\eta_2,\quad
\tau_t=2i\lambda(\eta_{1}^{2}-\eta_{2}^2)-2\eta_{1x}\eta_{2x},
\]
we find the potential
\begin{eqnarray}
\tau&=&{i\over
\sqrt{\kappa^2-c^2}}\Biggl[12\kappa\left[c_1c_2(c+\kappa)+c_3c_4(\kappa-c)\right]t
+2\left[c_1c_2(c+\kappa)-c_3c_4(\kappa-c)\right]x\nonumber\\
&&+{c_1^2(c+\kappa)\over 2\sqrt{\kappa}} e^{2\sqrt{\kappa}(2\kappa
t +x)} -{c_2^2(c+\kappa)\over 2\sqrt{\kappa}}
e^{-2\sqrt{\kappa}(2\kappa t+x)} +{ic_3^2(c-\kappa)\over
2\sqrt{\kappa}} e^{-2i\sqrt{\kappa}(-2\kappa t+x)}\nonumber\\
&& - {ic_4^2(c-\kappa)\over 2\sqrt{\kappa}}
e^{2i\sqrt{\kappa}(-2\kappa t
+x)}-\frac{(1+i)c_2c_4c}{\sqrt{\kappa}}e^{-\sqrt{\kappa}
[2(1+i)\kappa t+(1-i)x]}\nonumber\\
&&-\frac{(1-i)c_2c_3c}{\sqrt{\kappa}}e^{-\sqrt{\kappa}
[2(1-i)\kappa t+(1+i)x]}
+\frac{(1+i)c_1c_3c}{\sqrt{\kappa}}e^{\sqrt{\kappa}
[2(1+i)\kappa t+(1-i)x]}\nonumber\\
&&\left.+\frac{(1-i)c_1c_4c}{\sqrt{\kappa}}e^{\sqrt{\kappa}[2(1-i)\kappa
t +(1+i)x]}
 \right]\label{p2}
\end{eqnarray}
and the solutions for (\ref{hs}) are
\[
u=(\ln{\tau})_{xx}, \qquad
v=c+{\eta_{1x}\eta_2-\eta_1\eta_{2x}\over \tau},
\]
where $\eta_1$, $\eta_2$ and $\tau$ are given by (\ref{ex21}),
(\ref{ex22}) and (\ref{p2}) respectively. Our solutions contain
the known $1$-soliton solutions as special cases. Indeed, by
setting $c_1=c_3=0$, we arrive at
\begin{eqnarray*}
\tau&=&-\frac{ic_2^2}{2}\sqrt{{(c+\kappa)\over \kappa(\kappa-c)}}
e^{-2\sqrt{\kappa}(2\kappa
t+x)}\left[1+\frac{2(1+i)cc_4}{c_2(c+\kappa)}e^\theta+\frac{ic_4^2(c-\kappa)}{c_2^2(c+\kappa)}e^{2\theta}\right],\\
\eta_{1x}\eta_2-\eta_1\eta_{2x}&=&\frac{2(i-1)\kappa^{3\over
2}c_2c_4}{\sqrt{\kappa^2-c^2}}e^{-2\sqrt{\kappa}(2\kappa t+x)
}e^\theta
\end{eqnarray*}
where $\theta=\sqrt{\kappa}[2(1-i)\kappa t+(1+i)x]$.
if we further put $c=0$, we obtain the $1$-soliton solution first
found by Hirota and Satsuma \cite{hs1}.
We remark that our DT allows us to
construct multi-soliton solutions for HS-KdV system (\ref{hs}).

\section{Conclusions} In this paper, we constructed a new DT for
the HS-KdV system. We notice that our DT is a binary DT while the
one constructed by Leble and Ustinov is a classical DT. The
situation is similar to the celebrated KdV case. We obtained our
DT by using two methods, namely Painlev\'{e} analysis and
reduction method. In general, one may obtain a B\"{a}cklund
transformation if the wave functions involved in a DT are
eliminated. Thus, one may ask if it is possible to obtain  a
B\"{a}cklund transformation from our new DT. We realize that it is
not trivial to do this sort of elimination in the present case. On
the other hand, we remark that DT is easier to handle than
B\"{a}cklund transformation in order to construct explicit
solutions.

\bigskip

{\bf Acknowledgements}.  It is our pleasure to thank International
Centre for Theoretical Physics for hospitality.  The work is
supported in part by Ministry of Education of China and National
Natural Science Foundation of China (grant number 1997194).


\begin{thebibliography}{99}

\bibitem{hs1} Hirota R and Satsuma J 1981 {\em Phys. Lett. A} {\bf 85} 407.

\bibitem{hs2} Hirota R and Satsuma J 1982 {\em J. Phys. Soc. of Japan} {\bf
51}, 3390.

\bibitem{df} Dodd R and Fordy A P 1982 {\em Phys. Lett. A} {\bf 89} 168.

\bibitem{w} Wilson G 1982 {\em Phys. Lett. A} {\bf 89} 332.

\bibitem{le} Levi D 1983  {\em Phys. Lett. A} {\bf 95} 7.

\bibitem{ta} Tam W H, Ma W X, Hu X B and Wang D L 2000 {\em J. Phys. Soc.
Japan} {\bf 69} 45.

\bibitem{lu} Leble S B and Ustinov N V 1993 {\em J. Phys. A: Math. Gen.} {\bf
26 } 5007.

\bibitem{li} Li Y 1992  {\em Sci. China Ser. A} {\bf 35} 1050.

\bibitem{mu} Musette M 1999 in {\em Painlev\'{e} Property: one century later}, edited by R
Conte (Springer-Verlag).

\bibitem{eg} Estevez P G, Conde E and Gordoa P R 1998 {\em J. Nonlinear Math.
Phys.} {\bf 5} 82.

\bibitem{w1} Weiss J 1984 {\em J. Math, Phys.} {\bf 25} 2226.

\bibitem{w2} Weiss J 1985 {\em J. Math. Phys.} {\bf 26} 2174.

\bibitem{ni} Nimmo J J C 1995  in {\em Nonlinear Evolution Equations and
Dynamical Systems, NEEDS '94}, edited by V G Makhankov, A R Bishop and D
D Holm (World Scientific).

\end{thebibliography}
\end{document}